\documentclass[a4paper,10pt,twocolumn]{article}

\title{Solving Partial Differential Equations with Monte Carlo / Random Walk on an Analog-Digital Hybrid Computer}

\usepackage{amsmath}
\usepackage{amsfonts}
\usepackage{amssymb}
\usepackage{graphicx}
\usepackage{xcolor}
\usepackage[Option]{overpic}
\usepackage{todonotes}
\usepackage{multicol}
\usepackage{tikz}
\usepackage{algpseudocode} 
\usepackage{algorithm} 
\usepackage{titling}

\usepackage[T1]{fontenc}
\usepackage[nooneline,bf]{caption} 
\usepackage{times}
\usepackage{multicol}
\usepackage{amsmath}
\usepackage{amssymb}

\usepackage[hyperindex=true,hidelinks,pdfusetitle]{hyperref}
\def\BibTeX{{\rm B\kern-.05em{\sc i\kern-.025em b}\kern-.08em
    T\kern-.1667em\lower.7ex\hbox{E}\kern-.125emX}}

\small

\newenvironment*{mytitle}{\begin{LARGE}\bf}{\end{LARGE}\\[1.5ex]}%
\newenvironment*{myabstract}{\begin{Large}\bf}{\end{Large}\\[2.5ex]}%

\makeatletter
{}

\makeatother

\begin{document}

\newcommand{\anabrid}[1]{\, anabrid GmbH, \, #1@anabrid.com, Germany}
\twocolumn[
 \begin{@twocolumnfalse}
	
\begin{mytitle} Solving Partial Differential Equations with Monte Carlo / Random Walk on an Analog-Digital Hybrid Computer\end{mytitle}

Dirk Killat, anabrid GmbH and Brandenburg University of Technology,   killat@b-tu.de, Germany \\
Sven K\"oppel, \anabrid{koeppel} \\
Bernd Ulmann, anabrid GmbH and FOM University of Applied Sciences,  bernd.ulmann@fom.de, Germany \\
Lucas Wetzel, \anabrid{wetzel} \\

\vspace{5ex} 

\begin{myabstract} Abstract \end{myabstract}
 Current digital computers are about to hit basic physical boundaries with respect to integration density, clock frequencies, and particularly energy consumption. This requires the application of new computing paradigms, such as quantum and analog computing in the near future. Although neither quantum nor analog computer are general purpose computers they will play an important role as co-processors to offload certain classes of compute intensive tasks from classic digital computers, thereby not only reducing run time but also and foremost power consumption. 

 In this work, we describe a random walk approach to the solution of certain types of partial differential equations which is well suited for combinations of digital and analog computers (hybrid computers). The experiments were performed on an Analog Paradigm Model-1 analog computer attached to a digital computer by means of a hybrid interface. At the end we give some estimates of speedups and power consumption obtainable by using future analog computers on chip.

\vspace{4ex}	

\end{@twocolumnfalse}
]

\section{Introduction}

The prospect that -- extrapolating at the current growth rates -- the energy required to support the global computational demands will exceed the available resources within the next few decades \cite{cloudcoal,Andrae2015} highlights the need for more energy efficient approaches to computation. Non-traditional computing architectures (also refered to as \emph{unconventional} or \emph{exotic} computing) are about to close the gap between computational needs and performance delivered by existing digital architectures. Amongst them, there are for instance natural, neuromorphic or quantum computing approaches
\cite{Calude99aglimpse,Schuman.5192017,Ziegler2020,Zhou_2020,Georgescu_2014,Kenden2010}. In particular for data-heavy applications such as AI, novel materials and \emph{In-Memory Computing} are worked on \cite{Ielmini2020,Sebastian2020,Bavikadi2020}. A different approach are analog and mixed analog-digital computers as promising candidates to deliver high computational output at low energy demand for certain fields of applications \cite{Bournez2018,MacLennan2004,MacLennan2012,MacLennan2019}. In this respect, the most important properties of analog computers are the fully parallel computation and the high energy efficiency of such machines. {This comes at a cost as the precision of operations is basically limited to a signal-to-noise ratio of about $60$\,dB, also contributing to the high energy efficiency due to \textsc{Landauer}'s principle} \cite{LandauerFrank}.

In order to perform a computation, analog architectures make use of transferring a problem task into an analogue problem that can be implemented within the structure of the analog computer, e.\,g., an electrical circuit. The results of the analog computation, i.\,e., the system's continuous state variables, are then measured yielding the desired results which can be stored and post processed on a digital computer. Such a hybrid architecture (the combination of a digital computer with an analog computer) therefore exploits the advantages of both concepts -- analog and digital processing -- to perform computations fast and efficiently. Classic analog computers feature a variety of computing elements, including integrators with time as the free variable. Using modern technologies these ideas can be extended considerably, yielding techniques such as in-memory computing.

Given the continuous value nature of analog architectures, they are ideally suited for simulating or tracking fast time-evolutions of dynamical systems such as for instance in artificial intelligence (\emph{AI}) but also the broad class of partial differential equations (\emph{PDE}s), which are of central interest for describing problems in science and industry. In \cite{koppel2021_UsingAnaloga,krause} we have previously proposed the application of analog computers for fluid dynamics and molecular dynamics, both readily described by PDEs. In this work, we concentrate on stochastic differential equations (\emph{SDE}s) and the \textsc{Feynman}-\textsc{Kac} approach \cite{doi:10.1142/9789812567635_0001,Kac1949}.

In this work we present a proof of principle that parabolic PDEs, such as e.g., the heat equation, can be solved on hybrid computers using a Monte Carlo/Random Walk (\emph{MC/RW}) approach. This has been implemented on a modern modular analog computer, the \emph{Analog Paradigm Model-1} computer \cite{Model1Handbook,ap2} which is programmed in a classic fashion using patch cables to connect computing elements. Using power consumption and time to solution of this setup as a reference, we evaluate how efficiently such computations can be carried out after optimization. We also consider briefly future developments which will lead to reconfigurable analog computers on chip in CMOS technology, tightly integrated with a digital computer. This will considerably reduce the overall power consumption and overheads associated with the communication between the analog and digital parts of the system. The results are compared to the energy consumption of modern digital computers and are discussed in terms of their implications for next generation of microelectronic computation devices.

  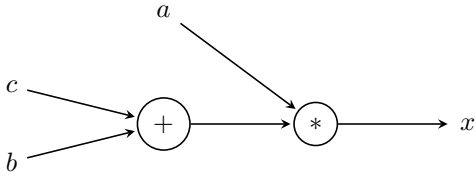
\begin{figure}
	\centering
	\begin{tikzpicture}[
		> = stealth, 
		shorten > = 1pt, 
		auto,
		node distance = 3cm, 
		semithick 
		]
		
		\tikzstyle{every state}=[
		draw = black,
		thick,
		fill = white,
		minimum size = 4mm
		]
		
		\node (B) at (0, 0) {$b$};
		\node (C) at (0, 1) {$c$};
		\node (PLUS) at (2, 0.5) [circle, draw] {$+$};
		\draw[->] (B) to (PLUS);
		\draw[->] (C) to (PLUS);
		\node (MULT) at (4, .5) [circle, draw] {$*$};
		\draw[->] (PLUS) to (MULT);
		\node (A) at (2, 2) {$a$};
		\draw [->] (A) to (MULT);
		\node (RESULT) at (6, .5) {$x$};
		\draw[->] (MULT) to (RESULT);
	\end{tikzpicture}
	\caption{Analog computer setup (circuit) for computing $x=a(b+c)$}
	\label{pic_ac_abc}
\end{figure}

\subsection{Basics of analog and hybrid computers}
 Classic stored program digital computers are machines capable of executing algorithms in a step-by-step fashion with individual instructions (and, in the case of a von \textsc{Neumann} machine also data) stored in some main memory. As powerful as this approach is, it forces the machine to a mainly sequential way of operation. 

 The following example shows the basic characteristics of this approach quite clearly. The expression $x=a(b+c)$ is to be solved for given values of $a$, $b$, and $c$, which are stored in memory. To compute the result $x$, three load operations are required to load the values into processor registers. This would not be necessary in case of a machine implementing instructions capable of working on values stored in memory directly. Nevertheless, the memory accesses would be still there although not as explicit instructions but disguised as an addition and multiplication operation. Then the value of $x$ can be computed by executing an addition and a multiplication. Storing $x$ back into memory would complete this little program. All in all this problem would require six individual instructions. These will be executed in a somewhat overlapping fashion by most modern digital computers but real parallelism would be hard to achieve.
 
  An analog computer is based on a completely different computational paradigm to a digital computer, as it does not use an algorithm in the classic sense and even has no memory at all. Instead, an analog computer consists of a plethora of computing elements, each capable of performing a basic mathematical operation such as summation, multiplication, or integration with time as the free variable. An analog computer program specifies how these computing elements are to be connected to each other in order to form an analog of the problem under consideration. It thus is a directed graph with computing elements at its nodes and vertices between these \cite{ap2}. Variables in such a setup are typically represented by (continuous) voltages or currents instead of bit sequences, vastly simplifying the connections between computing elements. 

  The above example of computing $x=a(b+c)$ would be solved on an analog computer as shown in Figure \ref{pic_ac_abc}. This program requires two computing elements, one summer and one multiplier as well as five vertices connecting these devices with their respective input values, etc. 
  
 This approach has a number of advantages over classic digital computers. First of all, all computing elements work in full parallelism with no need for memory accesses at all, no need for synchronisation, etc. In addition to this analog computers are highly energy efficient as long as limited precision is acceptable in the results obtained. The actual symbols for analog computing elements are different from this shown in this picture and will be explained below when required.

 The analog computer used in this study was an \emph{Analog Paradigm Model-1} computer manufactured by \emph{anabrid GmbH}. This is a recent modular analog computer and features integrators, summers, multipliers, comparators, and a hybrid computer interface which allows it to be coupled to a digital computer for parametrisation, control, and data aquisition. The computer uses physical voltages in the interval $[-10,10]\,\text{V}$ to represent values which are mapped onto a logical number representation over the domain $[-1,+1]$ with a precision of about $10^{-4}$. Any computation which yields a number outside of this domain results in an \emph{overload}. Given the domain, relative as well as absolute errors with respect to the maximum number representable are the same.

\section{Random Walks for Solving PDEs}
PDEs are amongst the most important mathematical frameworks in science and engineering. Despite their descriptive power, almost all non-trivial problems are not analytically solvable but require simplifications and approximations instead. Today, numerical methods dominate the solution strategies. They can be classified by a variety of properties. One distinguishes for instance between grid-based methods such as finite difference/volume/element methods (\emph{FD/FV/FEM}) or grid-free methods such as spectral methods \cite{Boyd2001} or stochastic attempts. Another differentiation is the applicability on PDE problem classes. One central property is the sign of the characteristics, indicating an elliptic, parabolic or hyperbolic problem. Where hyperbolic problems typically describe causal phenomena in physics undergoing some time-evolution, elliptic and parabolic systems typically describe stationary processes with no intrinsic information propagation direction. Therefore, a solution to an elliptic system is often the solution to an optimization problem. This work will focus on elliptic and parabolic systems.

Therefore, we revisit the \textsc{Feynman}-\textsc{Kac} method, which establishes a mapping of a partial differential equation onto the expectation value of an associated SDE. Accordingly, many realizations of the stochastic differential equation need to be computed to obtain the expected value. This task of evolving the stochastic process in time can be implemented on an analog computer while the computation of the expected value is then delegated to an attached digital computer. Using electronic noise sources also makes it possible to avoid complex pseudo-random number algorithms.

The Monte Carlo/Random Walk method can solve a specific set of $2^\text{nd}$ order PDEs and can be implemented on digital-analog hybrid setups. This is a grid free method which can handle complex domain geometries and is able to provide a solution at any point without the requirement of solving over the whole domain. For a detailed summary on MC methods see \cite{sawhney2022_GridfreeMonte,sawhney2020_MonteCarlo,Milewski2019}. For a general introduction into SDEs, see \cite{oksendal2013_StochasticDifferential} whereas \cite{2020_StochasticProcesses} provides an outline on using stochastic processes for boundary value problems (\emph{BVP}).
%

Spectral methods basically translate the PDE to be solved into a linear algebraic problem, thus allowing the whole system to be solved using standard techniques (assuming that sufficient system resources are available). Another approach are meshless methods, an example of which is shown in the following \cite{Milewski2018}. A particular useful example is the \textsc{Feynman}-\textsc{Kac} method. It translates the PDE so that it can be solved with a stochastic process. The main idea is to trace back the solution within a spatial domain from the boundary by carrying out random walks starting at a certain initial position, eventually hitting some boundary coordinate.
\subsection{Feynman-Kac}
At the heart of this technique is the formula
\begin{equation}
    \partial_t u + \omega \partial_x u + \alpha \partial_x^2 u - \sigma u + f = 0.
\end{equation}

This system describes a subclass of parabolic problems which has a number of interesting special cases, like the heat equation, the \textsc{Black}-\textsc{Scholes} model, the \textsc{Schrödinger} equation, \textsc{Fokker}-\textsc{Planck}, \textsc{Hamilton}-\textsc{Jacobi} and \textsc{Ornstein}-\textsc{Uhlenbeck} equations \cite{Alghassi:2021iod}, \cite[pp.~108~ff.]{Evans2013}. Here, the unknown $u$ and the parameters $\mu, \sigma, V, f$ are all fields like $\sigma=\sigma(t,x)$ in one spatial and one temporal dimension.

Here we will focus on the dimension-agnostic form for a stationary elliptic problem, i.\,e. $\partial_t u=0$, so that
\begin{equation}
   \nabla \cdot (\alpha \nabla u)+\vec{\omega}\cdot\nabla u -\sigma u + f = 0 \,,\label{eq:genPDE}
\end{equation}
with differential operators Nabla $\nabla_i = \partial_i$ and \textsc{Laplacian} $\Delta = \nabla^2$. The concept for solving the \textsc{Laplacian} $\Delta u=0$ boundary value problem by sampling the domain with a \textsc{Brownian} motion (\textsc{Wiener} process) was pioneered by \textsc{Kakutani}  \cite{kakutani1944_TwodimensionalBrownian}. This concept can be readily extended to \eqref{eq:genPDE}. The main idea is to define a stochastical differential equation for the given PDE by an \textsc{Itô}-diffusion
\begin{align}
  dX_t = \mu(X,t)dt + \sigma(X,t)dW_t,
\end{align}
where $\mu(X,t)$ and $\sigma(X,t)$ are functions and $W_t$ a \textsc{Wiener} process. The exit time for such a process is defined as:
\begin{align}
  \tau := \inf\left\{t \geq 0 | X(t) \notin \Omega \right\}. 
\end{align}
The algorithmic approach uses this exit time for an estimation of the fields
value within the domain,
\begin{equation}
  u(\vec x) = e^{-\sigma \tau} u(\vec x'),
  \quad \text{with} ~\vec x \in \Omega, ~ \vec x' \in \partial\,\Omega \,.
  \label{eq:exp-weight}
\end{equation}

A random walk is typically defined as a discrete stepwise process, while diffusion is understood as a continuous process. When tracking a random process naively checking the exit condition repeatedly at discrete time intervals, the exit time will always be overestimated. In the limit, with time steps going to zero, this overestimate will also approach zero.
\subsection{Test problem: Laplacian}\label{def:benchmark}
As a benchmark problem we consider the problem of finding solutions for a simple PDE using an analog computer coupled with a digital computer and compare these with a purely digital approach. 
The \textsc{Laplace} equation is an ideal test candidate: Solutions to symmetric geometries can be found analytically, more complex ones by means of \textsc{Fourier} series or approaches based on \textsc{Green}'s functions. Furthermore, many numerical methods exist for approximate approaches. The extension to the heat equation is given in a straightforward fashion by extending the differential operator $\Delta \to (\partial_t^2 + \Delta)$.
Non-homogenuous source terms enter on the right hand side as in $\Delta u = s$ and are only present on the boundary in the following scenario (eq. \ref{eq:boundary-values}).

The benchmark scenario sketched in the following is described by the two-dimensional spatial domain
\begin{align}
    \Theta_\pm &= \left\{~ \vec x \, |  \,
      \sqrt{(x_0 \pm 0.35)^2 + (y \mp 0.35)^2} < 0.25
      ~\right\} \nonumber \\
    \Xi    &= [-1,+1]^2 \nonumber \\
    \Omega &= \Xi \setminus \left( \Theta_+ ~\cup~ \Theta_- \right)
\end{align}
over the real numbers. That is, a square domain with two enclosed circles. Thus, there are three
distinct boundaries: The square one of the outer simulation domain and those of the two circles. The boundary values at $u(\vec x, t)\in \partial \, \Omega$ are defined as
\begin{equation}
    u(\vec x, t) =
    \begin{cases}
       0  & \text{if}~ \vec x \in \partial\, \Xi \\
       -1 & \text{if}~ \vec x \in \partial\, \Theta_+  \\
       +1 & \text{if}~ \vec x \in \partial\,  \Theta_-  \\
    \end{cases}
    \label{eq:boundary-values}
\end{equation}

\begin{figure}
  \centering
  \includegraphics[width=\columnwidth]{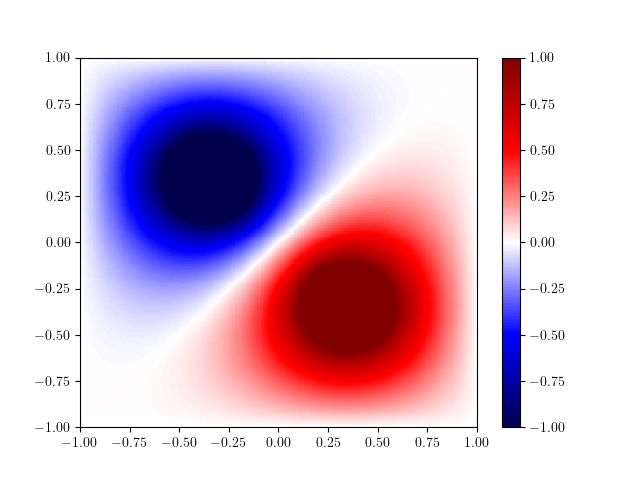} 
  \caption{Color-encoded plot of the solution obtained with a finite differences approach on the $200\times200$ grid executed on a digital computer. The full simulation domain including the two circular cut-outs in the upper left and lower right quadrant is shown.}
  \label{fig:digital-fd}
%
\vfill
  \centering
  \includegraphics[width=\columnwidth]{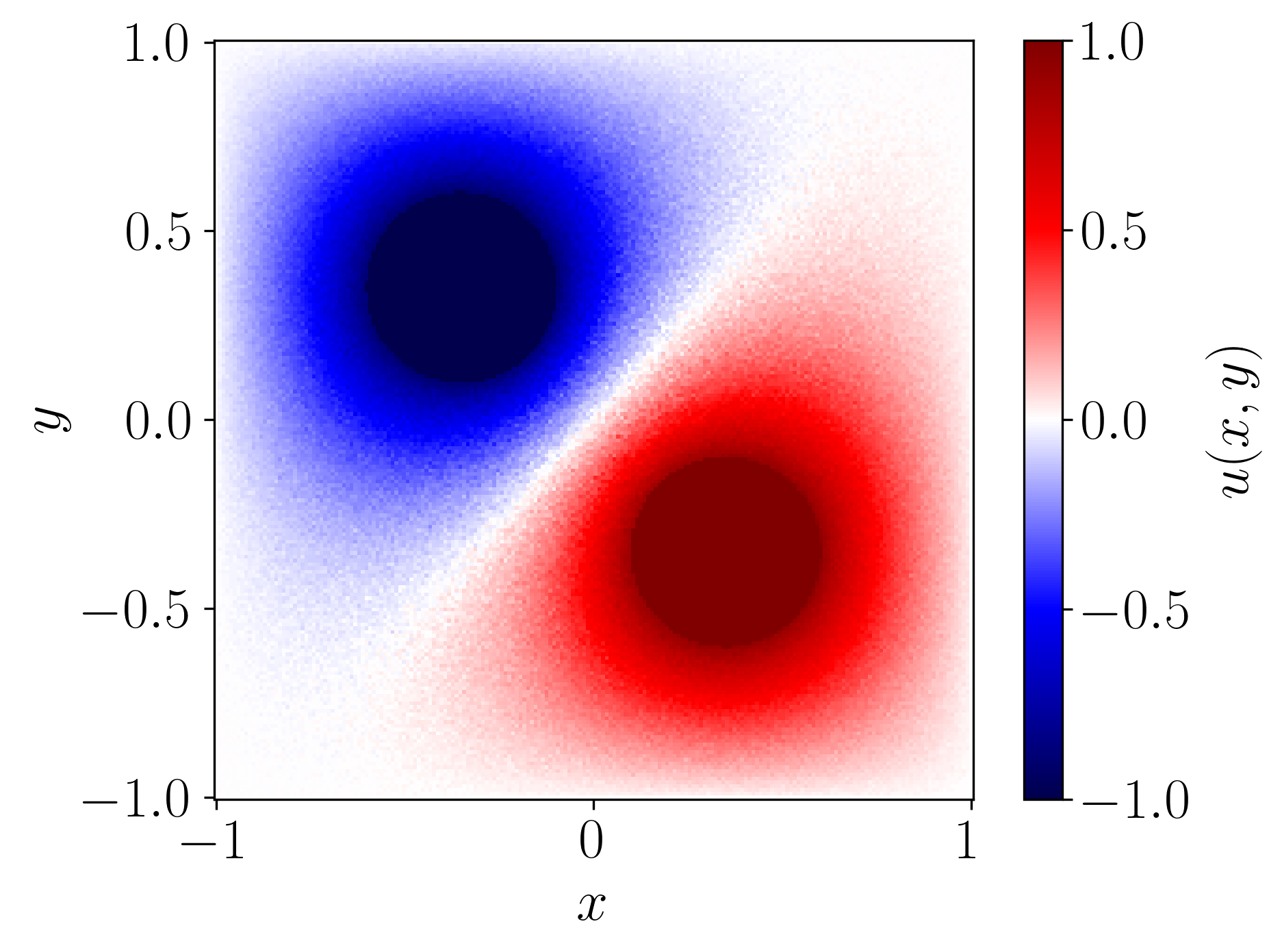} 
  \caption{Color-encoded plot showing the result of a 49\,h analog Monte Carlo/Random Walk approach with 800 runs per starting point $(x,y)$. The result obtained is well matched to that shown in Figure \ref{fig:digital-fd}. Note, however, the sandy fine-grained pattern caused by the stochastic approach.}
  \label{fig:analog-sol}
%
\vfill
  \centering
  \includegraphics[width=\columnwidth]{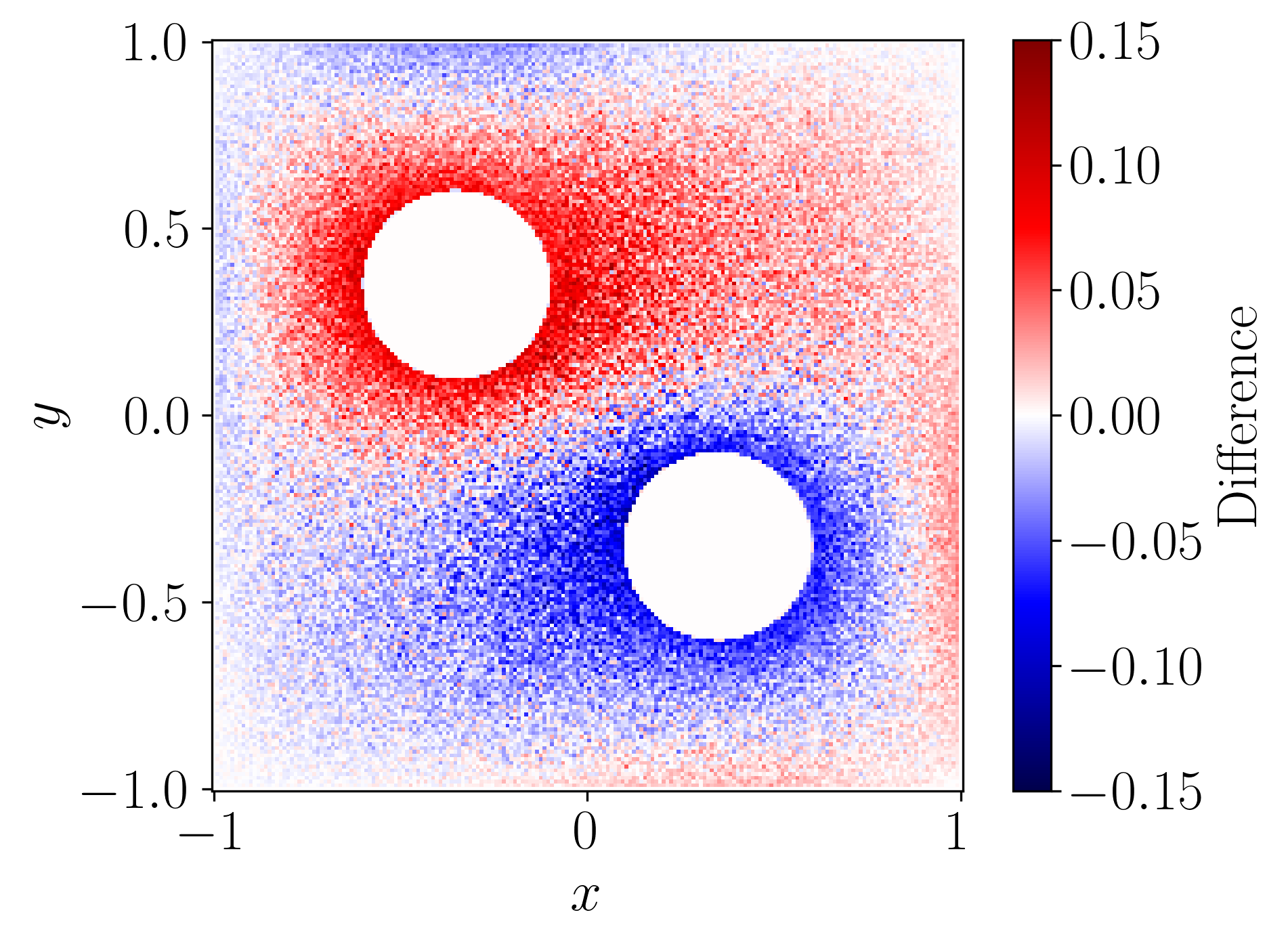} 
  \caption{Color-encoded plot of the absolute error between analog solution (Figure \ref{fig:analog-sol}) and more exact digital solution (Figure \ref{fig:digital-fd}). In this figure, the color code is different and shows maximum errors in the range of $\pm 15$\,\%.
  }
  \label{fig:erros}
\end{figure}

A near-to-exact solution of this setup is depicted in Figure~\ref{fig:digital-fd}.
\section{Implementation on a hybrid computer}
 The \textsc{Feynman}-\textsc{Kac} approach to solving PDEs is ideally suited for analog and hybrid computers and can be directly parallelised given that there are enough independent noise sources available. The basic idea is to implement the actual random walk on an analog computer with one noise source per dimension of the problem while the attached digital computer in a hybrid setup will do the necessary statistics over the individual random walks.  
 
\begin{figure}[t]
    \includegraphics[width=\columnwidth]{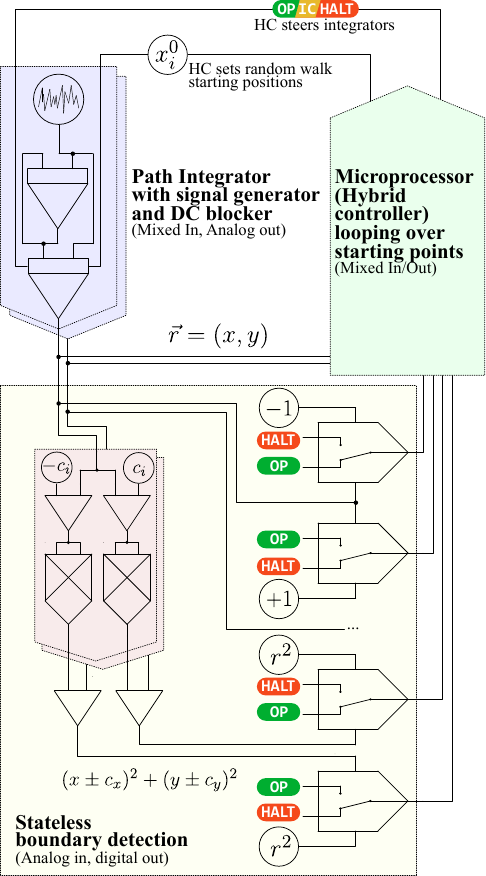}
    \caption{Block diagram/electronic circuit implementing a Monte Carlo/Random Walk solver for the heat equation. The diagram shows three large blocks, the stateful path integrator, the stateless boundary detection circuit, and the hybrid controller.
    }\label{fig:feynman-kac_analog-circuit}
\end{figure}

\subsection{The analog circuit}
 Figure \ref{fig:feynman-kac_analog-circuit} shows the setup for ths two-dimensional PDE using the \textsc{Feynman}-\textsc{Kac} technique. It consists of two basically identical circuits, one for each of the two dimensions, each fed an independent noise source. The noise signals used in this study were obtained by purely analog means, i.\,e., based on the noise of a PN-junction in a semiconductor with some signal and spectrum shaping applied. {The white noise generators used were of the type Wandel \& Goltermann RG-1 with 100kHz cutoff frequency.} These noise signals are fed to a circuit consisting of two integrators. An integrator is represented by a triangular shape with a rectangle on its input side. {Each integrator performs one integration with time as its free variable and performs an implicit change of sign which is due to its actual electronic implementation but of no relevance here.} The first integrator which has a (negative) feedback to itself generates a correction signal to remove any residual DC component of the input noise signal (cf. \cite[p.~80]{ap2}), while the integrator following it yields the position of the random walk for one dimension.

\begin{figure}[b!]
    \includegraphics[width=\columnwidth]{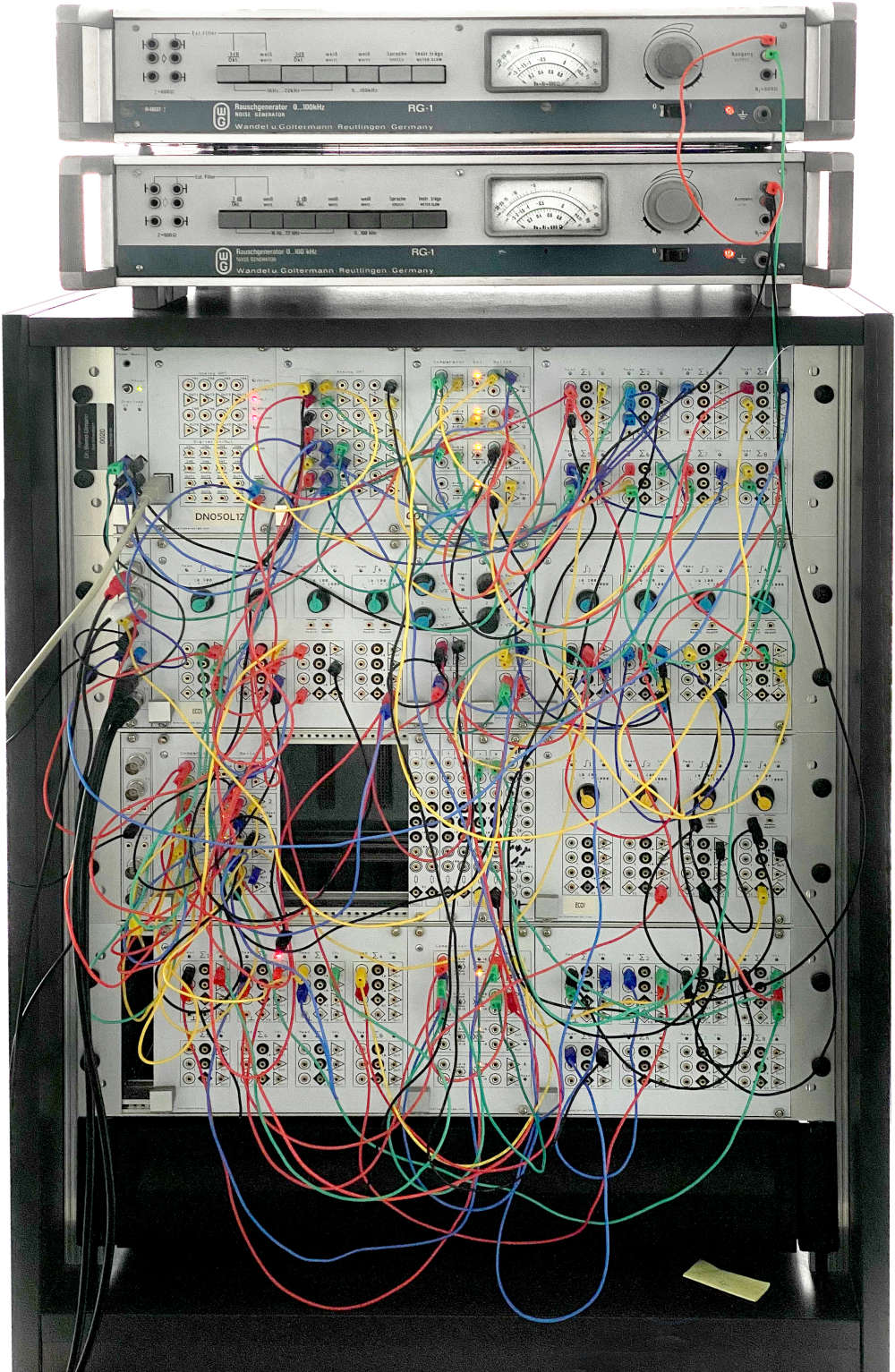}
    \caption{Photo of the experimental setup. The two noise generators sit on top of the rack mounted modular analog computer \emph{Analog Paradigm Model-1}. The hybrid controller is in the top left slot and can be recognised by the attached USB cable. The modules with  knobs are integrators, knobs set the time scaling constant for the integrators. Not shown are oscilloscopes for debugging and output as well as the standard laptop used for data post processing.}\label{img:photo-model1}
\end{figure}

 The resulting $x$- and $y$-components of the two-dimensional random walk are then fed to a circuit implementing the necessary boundary detection. This requires a number of summers, multipliers, and comparators \cite{ap2}.
 As soon as a particular random walk reaches a boundary, a HALT-signal is generated. This will halt the analog computation and signal the attached digital computer to read out the $x$ and $y$ values. Based on these the boundary value at this point is determined and taken into account to compute the expected value and thus the solution for the problem. In this example, the rectangular simulation domain contains two circles which are held at a constant temperature during simulation time. Figure \ref{img:photo-model1} shows the actual setup of the analog computer.
\begin{algorithm}[t]
	\begin{algorithmic}
		\State \textbf{enable} HALT on overload and external event
		\State \textbf{set} $c_x$, $c_y$, $r^2$
		\Comment{Initial simulation domain definition}
		\ForAll{$\vec x\in P_\Omega$}
		\Comment{Loop over domain grid points}
		\State $u\gets 0$
		\For{$1\leq j\leq N_t$} \Comment{Repetitions on $\vec x$ for statistics}
		\State \textbf{set} $x^0_i=x_i ~\forall~i \in [0,d-1]$
		\State \textbf{set OP}
		\Comment{Only HALT will terminate this}
		\If{HALT by Comparator}
		\State $g(x,y)\gets u(\vec x,t)$ \Comment{Implements Eq. \eqref{eq:boundary-values}}
		\Else
		\Comment{Halt by overload, treat as $\partial\Omega=\Xi$}
		\State $g(x,y)\gets 0$
		\EndIf
		\State $u\gets u+\text{e}^{-\sigma\tau}g(x,y)$
		\Comment{Impments Eq. \eqref{eq:exp-weight}}
		\EndFor
		\State $u(x,y)\gets{u}/{N_t}$
		\EndFor
		\State \textbf{plot} $u(x,y)$
	\end{algorithmic}
	\caption{MC/RW hybrid computer algorithm}\label{fig:algo}
\end{algorithm}

\subsection{The digital program}
The analog computer is tightly coupled to a digital computer by means of a hybrid controller (\emph{HC}). The digital computer executes the program as shown in Algorithm~\ref{fig:algo}.   
First, the hybrid controller is configured to stop the running analog computation when an external halt signal, generated by the boundary detection circuit, is applied. The central for-loop iterates over all points within the domain which are of interest. This encloses an inner loop that performs a number of individual runs. The loop body sets the initial conditions of the $x$- and $y$-integrators accordingly (these comprise the initial position of a particular random walk). When the random walk reaches a boundary, the analog computer is halted, and the digital computer reads the corresponding $x$- and $y$-values, determines the actual boundary condition at this point, and updates the expected value for this element in the domain.

This code can be implemented either fully on a microcontroller (MCU) or
distributed whereas the MCU only serves for data aquisition and some USB
uplinked desktop computer does the postprocessing, i.e. field value
reconstruction and subsequent plotting.


\subsection{Alternative boundary detection}
 Performing the actual random walk based on high-quality noise signals for each of the dimensions involved is simple while the purely analog detection of boundaries becomes quite a chore even for relatively simple shapes, as can be seen in Figure \ref{fig:feynman-kac_analog-circuit} -- most of the computing elements used in this setup are used for the secondary task of boundary detection.
 
\begin{figure}
  \centering
  \hspace*{-10mm}
  \begin{overpic}[width=.85\columnwidth,unit=1mm,tics=10,]{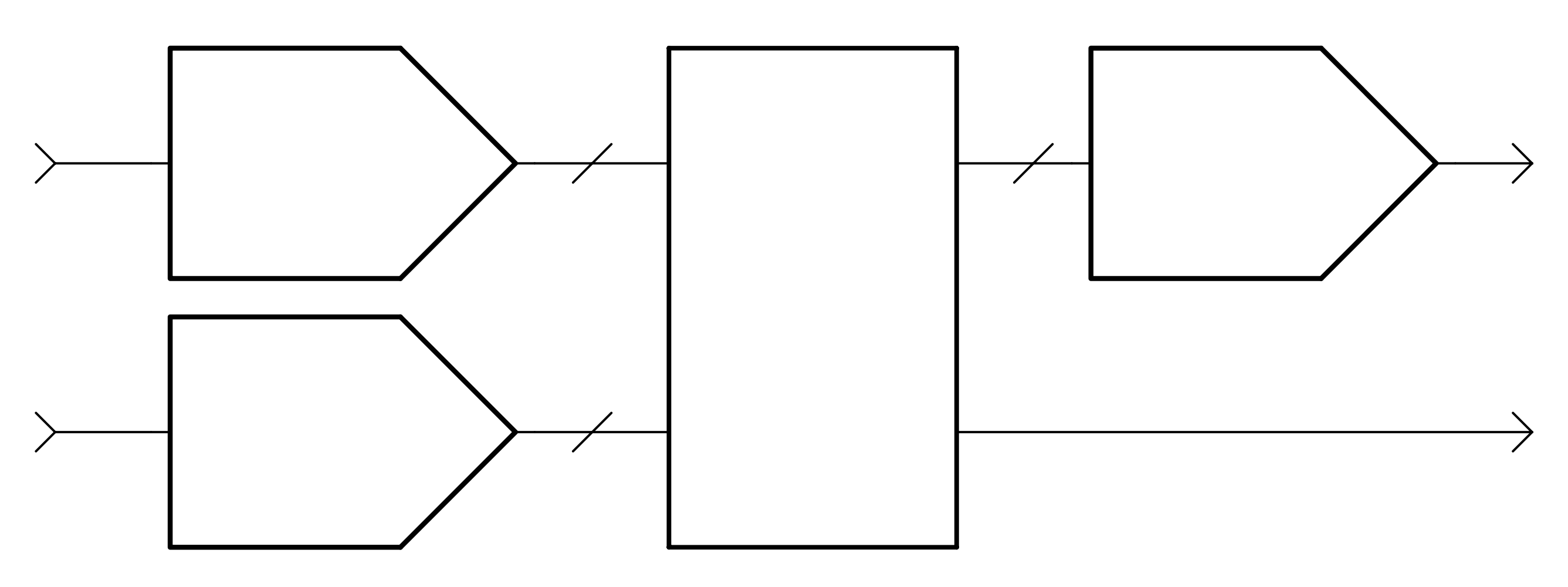} 
   \put(-2, 26){$x$}
   \put(-2, 9){$y$}
   \put(15, 25.5){ADC}
   \put(15, 8.5){ADC}
   \put(44, 18){\small Memory}  
   \put(74.5, 25.5){DAC}
   \put(100, 26){$f(x,y)$}
   \put(100, 9){$\chi(x,y)$}
  \end{overpic}
  \caption{Block diagram demonstrating the principle of operation of a table lookup function generator.}
  \label{pic_fg}
 \end{figure}
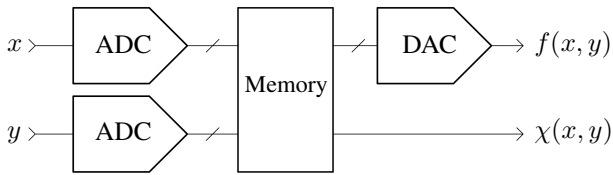

 A more generalised approach to this task could employ a function generator yielding two outputs, $f(x,y)$ representing the value at a certain point at a boundary, and characteristic function $\chi(x,y)$, a flag which will be set when $(x, y)$ is no longer inside the active region. The basic structure of such a function generator is shown in Figure \ref{pic_fg}. It is of the classic table-lookup type. Two analog-to-digital converters (ADCs) convert the continuous input signals $x$ and $y$ into suitable partial addresses for a (small) high-speed memory with a word length of $n$ bit. This feeds a digital-to-analog converter (DAC) with $n-1$ of its output bits, while the $n$-th bit is used for the characteristic function $\chi(x,y)$.
 
 A function generator like this could then be used as a generalised boundary detection circuit yielding a characteristic function $\chi(x,y)$ to halt the analog computation, notify the attached digital computer, and provide the boundary value $f(x,y)$. In addition to simplifying the overall setup, this would have the additional advantage that boundaries could be basically arbitrarily complex. Implementing a certain boundary would only involve writing suitable values to the lookup memory instead of designing a tailored analog circuit for this purpose. If a $256\times256$ grid with seven bit boundary values is sufficient, this would require two 8-bit ADCs, 64 kB of memory, and a seven-bit DAC. A function generator of this complexity can be easily implemented in CMOS technology and would allow for more generalised boundary shapes.
 
\section{Results}
The benchmark problem described in section \ref{def:benchmark} is solved on a reasonably dense grid of $N_x \times N_y = 200 \times 200$ points over the domain $[-1,1]^2$. $N_t=800$ individual random walks are executed per point to obtain precise expected values. Thus, $M := N_t \times N_x \times N_y \approx$ 33 million runs (realizations) are carried out. For the benchmark, the \emph{Analog Paradigm Model-1} is tested against a \emph{Intel\textsuperscript{\textcopyright} Whisky Lake} ``ultra-low power mobile'' processor (\emph{Core i7-8565U}) as a representative of a typical desktop-grade processor.
\subsection{Runtime and energy results}
The average time per run for a single realization of the random walk is $T^1_A \approx 5.4$\,ms on the \emph{Analog Paradigm Model-1} analog computer. This time does not take into account communication overheads for data acquisition such as the USB latency, which however is irrelevant given the unidirectional data flow between analog and digital computer. The serial approach (one random walk at a time) results in a total run time of about $T^M_A = 49$\,h (wall clock time).

The power requirement of the analog circuit is $P^A\approx3$\,W, where  20 computing elements are assumed with $150$\,mW average energy consumption \cite{koppel2021_UsingAnaloga}. To further simplify things, digital data aquisition etc. is not taken into account. This results approximately $E^A_M\approx 147$\,Wh of energy consumption for the analog computer.

\subsection{Interpretation and Discussion}
Run time and energy requirements of the digital and analog approaches are in the same ballpark. It should be noted that such benchmarks have, by nature, a large uncertainty, given by the complexity and large number of configuration options of the systems under consideration. For instance, a highly optimized serial code for the digital processor could easily achieve one order of magnitude better performance. Possible variations are discussed in section \ref{sec:wannabe-ic}.

It should be also noted that problems at this given size can be also solved efficiently and quickly with other solution methods. For instance, the PDE can be recast into a system of linear equations by means of finite differences over the a discretised solution domain. This results in a sparse matrix of size $10^4 \times 10^4$ with a density of $10^{-4}$. This system can be solved on the digital benchmark computer in a one-step approach within $T^M_{D2}=200$\,ms using a suitable numerical scheme. This solution time is three orders of magnitude smaller than the digital random walk, $T^M_{D2} \ll T^M_{D}$. However, the scaling of this matrix method is worse and in particular parallelisation can not be achieved. Typical matrix solvers scale like $T^M_{D2}=\mathcal{O}((N_xN_y)^{2.37 - 3})$ where the na\"ive serial runtime is $T^M_D=\mathcal{O}(M^2)$.
\section{Towards Integrated Circuitry}\label{sec:wannabe-ic}
In the previous section, we have shown that the discrete analog computer \emph{Model-1} shows performance comparable to a modern desktop for the given test problem. However, this basically compares 1970s level discrete analog electronics with 2020s level digital processor technology. In this section, we will show the route towards contemporary and future analog computer implementation and capabilities.

As a next step we already have a new analog computer, the \emph{Model-2}, running in a laboratory setup, which features $10$ times the bandwidth of a \emph{Model-1} system allowing for run times of  $T^{A2}_1\approx500\,\mu$sec per random walk with a similar power budget. This system, already being fully reconfigurable by the attached digital computer, has a much higher packing density (about ten times denser than a comparable \emph{Model-1}). This system is roughly halfway towards a highly integrated general-purpose analog computer on chip, called \emph{Model-3}. 

Based on \cite{koppel2021_UsingAnaloga} the \emph{Model-3} should exhibit $10^{2.5 \pm 0.5}$ times the bandwidth of the \emph{Model-1}. This would result in a run time of $T^{A3}_1\approx17\,\mu$sec per individual random walk and thus $T^{A3}_M\approx9$\,min for the 33 million realizations, without using parallel random walk. The power ratio between such an integrated circuit implementation and a \emph{Model-1} can be estimated as $P^A/P^{A3}=10^{-3}$, yielding $P^{A3}\approx3$\,mW, an energy consumption of $E^{A3}_1\approx51$\,nJ, and a total energy consumption of $E^{A3}_M\approx2$\,J.
\subsection{Parallelization}\label{sec:parallel}
The main advantage of the proposed PDE solution method is the elimination of communication between neighbouring points in the solution domain. This makes parallelization extremely easy as $n$ distinct analog random walk implementations yield a speedup of $n$. Compared with a \emph{Model-1} with its small number of computing elements, the more advanced \emph{Model-2} already offers some degree of possible speedup due to parallelization of individual random walks. The proposed chip, the \emph{Model-3}, will further increase this capacity. 

Depending on whether the boundary detection is carried out in software or hardware, we expect to be able to run between 20 to 100 random walks in parallel on a single microchip (assuming 65nm and roughly 10mm$^2$ of die area). This would yield $10^{3.4\pm0.4}$ random walks using 50 chips occupying roughly the physical volume of a \emph{Model-2} system, about $2\,300\,\text{cm}^3$). A super computer configuration consisting of $10^5$ chips would allow $10^{6.6\pm0.35}$ parallel random walks.

To put this into perspective, the Top500 supercomputer list \cite{Dongarra2011} is currently lead by the \emph{Frontier} system with $8,699,904\approx10^7$ cores and an overall power consumption of 22MW. In contrast to this, the 100,000 analog chips will only consume a few $10$\,kW and solve problems of this class about $T^D_1/T^{A3}_1 \approx150$ times as fast. This ratio does not even take into account slowdowns due to the digital communication overhead which are dominant in digital supercomputers of that size.
\section{Summary and Outlook}
Analog computers will be an integral part of tomorrow's computer systems to speed up the solution of certain types of problems while simultaneously reducing the overall energy consumption for such computations. This requires new approaches to computing due to the non-algorithmic nature of analog computers, an example of which has been demonstrated in this paper by solving partial differential equations using random walks. This approach is superior to algorithmic approaches when the problem size gets very large, in particular when only low precision solutions are required. 

The study can be extended in several ways: First, the analog-digital hybrid methods can be refined by implementing findings of the last decades in the community, such as guided random walks (for instance Hamiltonian Monte Carlo) or implementing ideas of quantum random walk approaches. Second, it very interesting to apply the method on a broader class of PDEs or on analog problems such as the probabilistic solution of very large linear equations. Third, software support may be improved to allow a broad use of the presented methods. This will require software libraries to tightly integrate the analog part into the digital domain. Fourth, the hardware can be improved considerably by integration, as presented in the theoretical estimates from \emph{Model $1\to2\to3$}. In future works, we want to underpin the theoretical findings with practical measurements on the discussed digital-analog computer architectures yet to built.

\subsubsection*{Acknowledgements}
We thank \textsc{Nick Baberuxki} for setup, run and analysis of the \emph{Model-1} experiments.
We thank \textsc{Maikel Hajiabadi} for the finite difference/LGS runs and analysis and literature
contributions.
We thank \textsc{Michael Steck} for contributions for more efficient Random Walk Microcontroller code.
The authors would like to thank Dr. \textsc{Chris Giles} for 
fruitful discussions and his meticulous proof reading.
\bibliographystyle{IEEEtran}
\bibliography{IEEEabrv,References}

\end{document}